\documentclass[aps,pra,twocolumn,superscriptaddress,showkeys,amsmath,amssymb]{revtex4-1}		
\usepackage{graphicx}
\usepackage{dcolumn}
\usepackage{bm}
\usepackage{hyperref}
\usepackage{subfig}

\usepackage[font=small,labelfont=bf,format=plain,justification=centerlast,labelsep=period]{caption}
\usepackage{txfonts}
\usepackage[usenames, dvipsnames]{color}

\usepackage{verbatim}
\usepackage[normalem]{ulem}

\begin{document}

\preprint{APS/123-QED}

\title{Large time-asymmetric quantum fluctuations in amplitude-intensity 
correlation \\measurements of a V-type three-level atom resonance 
fluorescence}

\author{L. Guti\'errez}
\affiliation{Centro de Investigaci\'on en Ingenier\'ia y Ciencias Aplicadas, 
Instituto de Investigaci\'on en Ciencias B\'asicas y Aplicadas,
Universidad Aut\'onoma del Estado de Morelos\\
Avenida Universidad 1001, 62209 Cuernavaca, Morelos, M\'exico}
\author{H. M. Castro-Beltr\'an}
\email{hcastro@uaem.mx}
\affiliation{Centro de Investigaci\'on en Ingenier\'ia y Ciencias Aplicadas, 
Instituto de Investigaci\'on en Ciencias B\'asicas y Aplicadas,
Universidad Aut\'onoma del Estado de Morelos\\
Avenida Universidad 1001, 62209 Cuernavaca, Morelos, M\'exico}
\author{R. Rom\'an-Ancheyta}
\affiliation{Instituto de Ciencias F\'isicas,
Universidad Nacional Aut\'onoma de M\'exico, Apartado Postal 48-3, 62251 Cuernavaca, Morelos, M\'exico}
\author{L. Horvath}
\affiliation{Department of Physics and Astronomy, Macquarie University, 
North Ryde, NSW, 2109, Sydney, Australia}
%

\begin{abstract}
In this paper we show that the scattered field of a bichromatically driven 
V-type three-level atom exhibit asymmetry and large violation of classical 
bounds in amplitude-intensity correlations. These features result from the 
noncommutativity of amplitude and intensity field operators, and the 
strong non-Gaussian fluctuations in this system. The amplitude-intensity 
correlations of resonance fluorescence, with its large third-order 
fluctuations, describe the nonclassical features of the emitted field more 
accurately than the second-order measure related to squeezing. Spectra 
and variances of these correlations, along intensity-intensity correlations, 
provide a wealth of supporting information. 

\begin{description}
\item[PACS number(s)]
42.50.Pq, 42.50.Lc, 42.50.Ct, 32.80.-t
\end{description}
\end{abstract}




\maketitle


\section{Introduction}  
The correlation among the intensity $I$ and a delayed quadrature amplitude 
$E_{\phi}$ of a quantum field, $\langle I(0) E_{\phi}(\tau) \rangle$, has been 
recently established as a genuine and powerful tool to study, observe, and 
identify quantum fluctuations of light. Given the conditional nature of this  measurement, it can reduce the issues of low quantum and collection 
efficiencies of detectors which affect weak squeezed light emitters, such as 
cavity QED and resonance fluorescence. There are two main approaches 
to the amplitude-intensity correlation (AIC), both being variants of the 
Hanbury-Brown-Twiss setup of intensity-intensity correlations \cite{HBT56}: 
One is conditional homodyne detection (CHD) 
\cite{CCFO00,FOCC00,CFO+04}, where the stop detector is replaced by a 
balanced homodyne detection setup; here, the AIC is explicitly and directly 
measured. In the other approach, homodyne correlation measurement 
\cite{Vogel91,Vogel95,KVM+17}, the input field to the Hanbury-Brown-Twiss 
setup consists of the source field mixed with a phase-selected reference 
field. In this case the AIC is one measured term and the variance is 
another; addition and subtraction of measurements for several phases, 
however, are necessary to extract the desired quadrature amplitude.   

Since the field's amplitude and intensity operators do not commute and 
have distinctive noise properties, time-asymmetric correlations, and hence 
non-Gaussian fluctuations (nonzero odd-order correlations) of the field, 
can be naturally detected by the AIC measurement scheme
\cite{DeCC02,MaCa08,CaGM14,XGJM15,XuMo15,WaFO16}. Indeed, 
asymmetric amplitude-intensity correlations were initially spotted in CHD 
simulations \cite{CCFO00,DeCC02} and experiments \cite{FOCC00} in 
cavity QED. Due to the relatively weak driving, the asymmetry observed 
in the correlation was small, meaning that the light fluctuations were 
approximately Gaussian. However, notoriously asymmetric and giant 
correlations have been predicted for the light scattered from the often 
ignored weak transition in a bichromatically driven V-type three-level atom 
(V3LA) \cite{MaCa08,CaGM14,XGJM15}. For the strong transition there 
is little deviation from the symmetry of the CHD correlation of a two-level 
atom \cite{hmcb10,CaGH15} or a single-laser driven 3LA with electron 
shelving \cite{CaRG16}, where the small Hilbert space inhibits the 
asymmetry. Large asymmetric correlations have also been predicted for 
a 3LA in the ladder configuration \cite{XGJM15}, for a pair of Rydberg 
atoms with blockade effect \cite{XuMo15}, and for a superconducting 
artificial atom \cite{WaFO16}. 

M\o lmer and coworkers have approached the asymmetry from the 
viewpoint of quantum measurement theory \cite{XGJM15,GaJM13}. 
They have demonstrated the value of working out the past state of a 
quantum system based on a photo-detection event in the present. In 
particular, they have described the properties of forward and backward 
time evolutions surrounding a photo-detection event and computed the 
amplitude-correlation function for a resonantly driven V3LA. It was clear 
from this work that the initial state of forward and backward time 
evolutions are very different because of the distinct initial conditions and 
because of the different steady states obtained from the forward and 
backward time evolution dynamics. M\o lmer and coworkers explained 
why the past quantum state is a better predictor of a photon counting 
event than the density matrix of the system alone.

In this paper, we investigate how the different fluctuations of the 
amplitude and intensity of the emitted field manifest in the asymmetry of 
the AIC. In particular, we employ this method within the CHD theory to 
study the role of the atomic level configuration and driving field conditions 
in a bichromatically driven single V3LA. We focus mainly on the weak 
transition, which displays asymmetry in a more striking way than the 
strong transition \cite{MaCa08}. To access the signature of non-Gaussian 
fluctuations we decompose the dipole field into average and noise terms 
to distill the third-order noise operator from the AIC. From this we explore 
the connection between AIC correlations and more typical measures of 
fluctuations such as spectra and variances. We use the third-order noise 
operator as a new tool to explore squeezing (or, more precisely, its 
deviation from it), the lack of detailed balance \cite{WCGS78}, and 
non-Gaussian atom-field fluctuations.

We organize this paper as follows: We outline the atom-laser system in 
Section 2. In Section 3 we consider intensity correlations of two photons 
from a single transition in a V3LA. In Section 4 we study the AIC by CHD 
and deal with the asymmetry, and second and third-order fluctuations. In 
Section 5 we consider measures of noise such as quadrature spectra 
and variance. Conclusions are given in Section 6 and a brief appendix 
contains additional analytical results.

\section{Atom-Laser Interaction}
\begin{figure}[t]
\includegraphics[width=8.5cm,height=5cm]{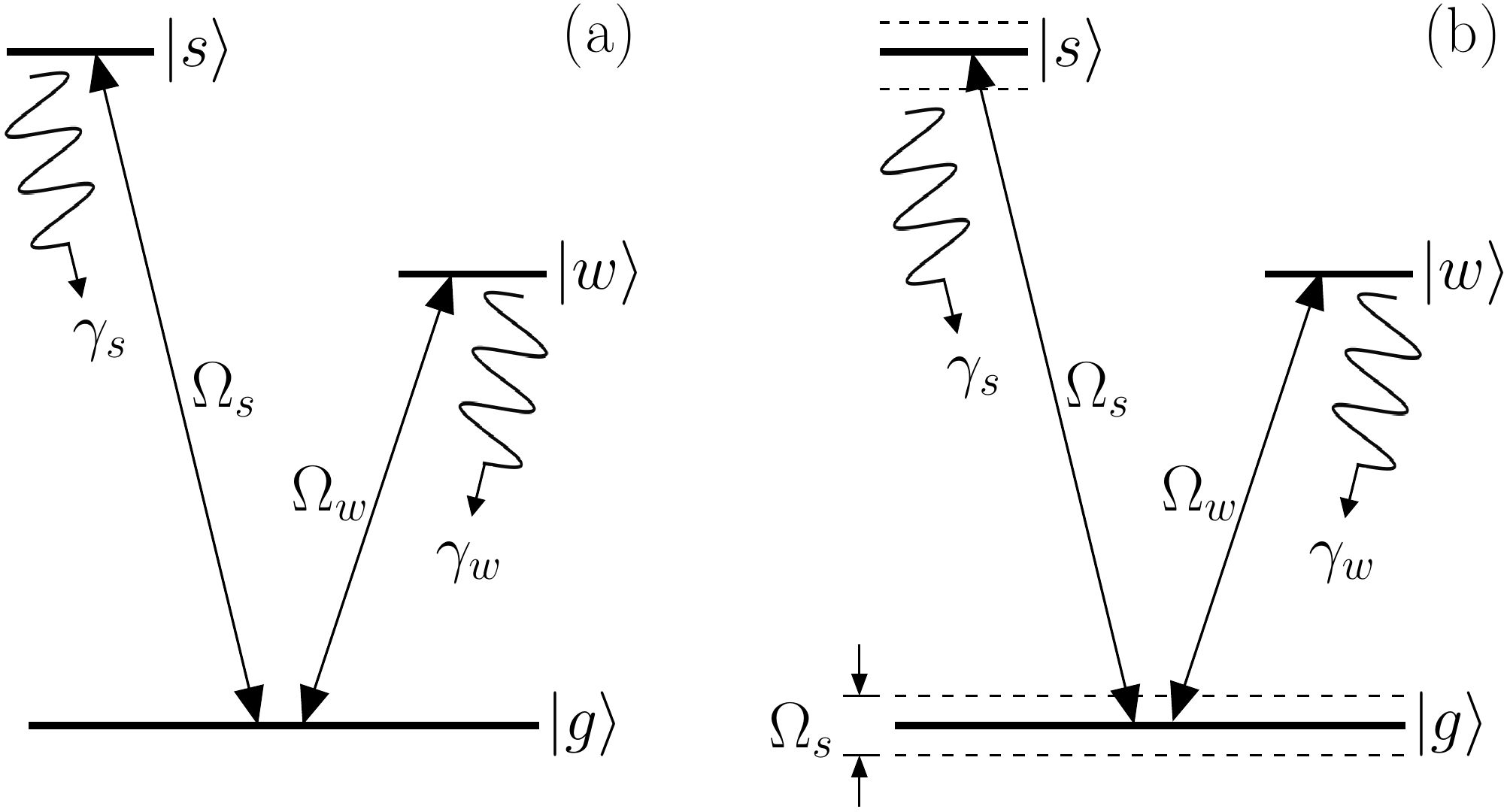}  
\caption{\label{fig:V3LA} 
(a) V-type three-level atom driven bichromatically (see main text for 
details). Since $\gamma_s > \gamma_w$ we call 
$|g \rangle - |s \rangle$ $\left( |g \rangle - |w \rangle \right)$ the strong 
(weak) transition. (b) If $\Omega_s \gg \gamma_s$ the strong transition 
experiences AC Stark splittings (dashed lines). }
\end{figure}

We consider a single V3LA with one ground state $|g \rangle$ coupled to 
two excited states $|s \rangle$ and $|w \rangle$ by monochromatic lasers 
with Rabi frequencies $\Omega_s$ and $\Omega_s$, see 
Fig.~\ref{fig:V3LA}(a). We assume that the transition frequencies are very 
different so that each laser couples two levels only. Hence, the transitions 
are coupled to independent reservoirs, thus neglecting effects of coherence 
among the excited states. In this configuration we can use broadband 
detectors that are able to distinguish light from the separate transitions. 
Two-time correlations for each of these decay channels have been 
calculated separately in this paper. 
%

This system is prototypical in studies of electron shelving \cite{PlKn97}, 
where only the light emitted in the strong dipole transition 
$|g \rangle - |s \rangle$ is detected; the light from the dipole-forbidden 
$|g \rangle - |w \rangle$ transition is so dilute that it would be buried in 
photo detection noise. In this paper, in contrast, we are interested in the 
light from a weaker  electric dipole transition $|g \rangle - |w \rangle$, 
where $\gamma_w$ is smaller than $\gamma_s$ by only one or two 
orders of magnitude, and excited above saturation to allow for a 
measurable fluorescence rate. For the strong transition we consider two 
excitation regimes: it is either excited (i) moderately, 
$\Omega_s = \gamma_s/2$, or (ii) strongly, $\Omega_s = 3.5 \gamma_s$. 
We consider a Rabi frequency of the weak transition, 
$\Omega_w \sim \gamma_w$, that is strong enough to compete with the 
strong transition in case (i), but acts as a probe in case (ii). In case (ii) the 
$|g \rangle - |s \rangle$ transition experiences an AC Stark splitting, see 
Fig.~\ref{fig:V3LA}(b), where the detuning from the weak transition 
$|g \rangle - |w \rangle$ takes the system into the Autler-Townes regime \cite{AuTo55,ZSAW00}.  

The master equation in a frame rotating at the laser frequencies is given 
by 
\begin{eqnarray} 
\dot{\rho} &=& -i \sum_{e=s,w} 
	\frac{\Omega_e}{2} \left[ \sigma_{eg} +\sigma_{ge}, \rho \right]
	-i\sum_{e=s,w} \Delta_e \left[ \sigma_{ee}, \rho \right]   \nonumber \\
&& +\sum_{e=s,w} \frac{\gamma_e}{2} ( 2\sigma_{ge} \rho \sigma_{eg} 
	- \sigma_{ee} \rho -\rho \sigma_{ee} ) 	\,,
\end{eqnarray}
where $\sigma_{jk} = |j \rangle\langle k|$ are Pauli pseudospin operators. 
For later reference, we define the steady state values of the atomic 
operators as $\alpha_{jk} = \langle \sigma_{jk} \rangle_{st}$. $\Delta_e$ 
are the laser detunings (which we set equal to zero through the rest of 
the paper). Due to the complexity of the parameter space, we refrain from 
trying analytical solutions, but it is not particularly difficult to extract 
conclusions from the observations.  

\section{Intensity-Intensity Correlation}  
We begin our investigation of the quantum fluctuations of the V3LA 
resonance fluorescence with a brief analysis of the intensity fluctuations, 
usually studied via the Hanbury-Brown-Twiss correlation, that is, the 
normalized probability of detection of two photons separated by a time 
delay $\tau$ \cite{CaWa76,KiMa76}, 
\begin{eqnarray} 	\label{eq:g2}
g_{ee}^{(2)}(\tau) = \frac{ \langle  \sigma_{eg}(0) \sigma_{eg}(\tau) 
	\sigma_{ge}(\tau) \sigma_{ge}(0)  \rangle }{ \alpha_{ee}^2 } \,.
\end{eqnarray}
For simplicity, we only consider the case where both photons come from 
the same transition, thus this correlation is intrinsically time-symmetric 
\cite{DeCC02}. For the V3LA this correlation has been considered in 
Ref.~\cite{PeLK86} in the electron shelving regime. We, as mentioned 
above, consider a less stringent situation for the weak transition. 
\begin{figure}[t]
 \includegraphics[width=8.5cm,height=7cm]{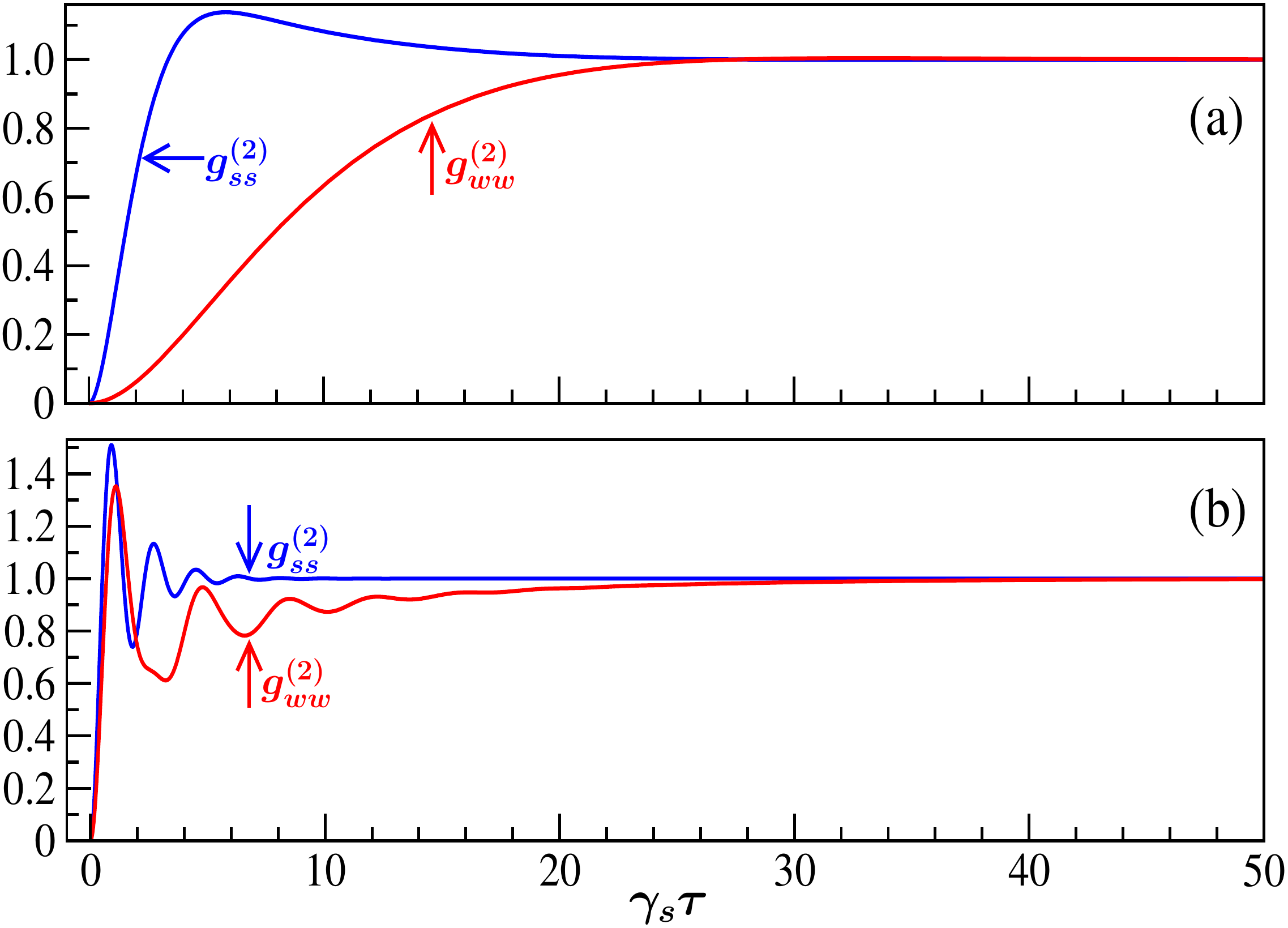}
\caption{\label{Fig_g2} 
Intensity-intensity correlations $g_{ss}^{(2)}(\tau)$ (blue) and 
$g_{ww}^{(2)}(\tau)$ (red). The parameters used are: 
$\gamma_w = \Omega_w =0.1 \gamma_s$, 
$\Delta_s = \Delta_w =0$, and (a) $\Omega_s =0.5 \gamma_s$, 
(b) $\Omega_s =3.5 \gamma_s$.}
\end{figure}

Figure~\ref{Fig_g2} shows intensity correlations when the Rabi frequency 
of the weak transition is fixed at a value above saturation, 
$\Omega_w = \gamma_w$, and the strong transition is driven either 
moderately, Fig.~\ref{Fig_g2}(a), or strongly, Fig.~\ref{Fig_g2}(b). For 
instance, there is the well-known nonclassical feature of antibunching: 
the atom cannot emit two photons simultaneously, $g^{(2)}(0) =0$  \cite{CaWa76,KiMa76}. More generally, $g^{(2)}(\tau) <1$ means that the 
atom is in the nonclassical state of antibunching. For the V3LA we observe 
this feature because the two transitions feed from the common ground 
state. However, there are notable differences in their evolution not 
observed in Ref. \cite{PeLK86}.  

In Fig.~\ref{Fig_g2}(a)  $\Omega_s/\gamma_s = 0.5$ and 
$\Omega_w/\gamma_w = 1$, that is, the weak transition is apparently 
more strongly driven but its net emission rate $\gamma_w \alpha_{ww}$ 
is smaller than that of the strong transition $\gamma_s \alpha_{ss}$. 
The strong transition (the one with larger $\gamma_s$) competes 
advantageously for transition probability with the weak transition; the  
average separation among photons in the strong transition is shorter than 
that for the weak transition. This explains the longer approach towards the 
value $g_{ww}^{(2)} = 1$, which characterizes independent photon 
emissions.  

In Fig.~\ref{Fig_g2}(b) we have $\Omega_s/\gamma_s = 3.5$ and 
$\Omega_w/\gamma_w = 1$. For the strong transition the transient period 
of $g_{ss}^{(2)}$ shows oscillations at a frequency near $\Omega_s$, 
damped at the approximate rate $3 \gamma_s/4$, just as it occurs for a 
two-level atom \cite{CaWa76,KiMa76}. The dressing of the strong 
transition, see Fig.~\ref{fig:V3LA}(b), forces the dynamics of the weak 
one to evolve at a frequency $\sim \Omega_s/2$. Interestingly and 
unusually, the oscillation regime of $g_{ww}^{(2)}$ occurs mostly 
\textit{below} unity, with a long term decay at the rate $\gamma_w$, 
more clearly manifested than in the case of $g_{ss}^{(2)}$. The weak 
transition is thus in a highly nonclassical state, which certainly calls for a 
deeper study, such as of its phase-dependent fluctuations.

\section{Amplitude-Intensity Correlation}  
\begin{figure}[t]
\includegraphics[width=8.5cm,height=6cm]{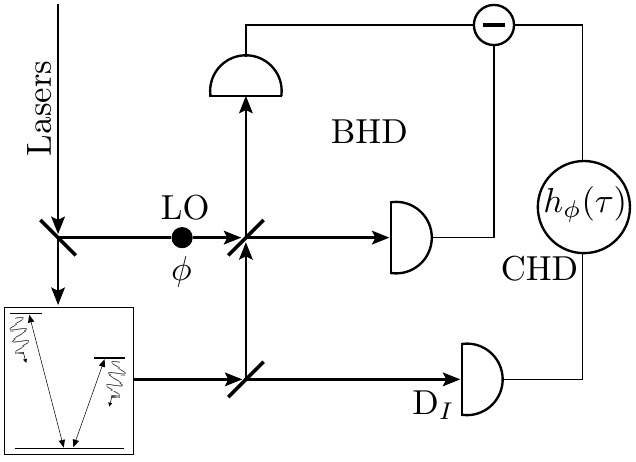}  
\vspace{2mm} 
\caption{\label{fig:chdsetup} 
Sketch of a conditional homodyne detection setup. See main text for 
details. The excitation is bichromatic but the local oscillator's frequency 
is only that of the observed transition.}
\end{figure}

We study the AIC via conditional homodyne detection, see 
Fig.~\ref{fig:chdsetup}. In this method, the amplitude of a quadrature of 
the emitted field, $E_{\phi} \propto \sigma_{\phi} = (\sigma_{eg} e^{-i\phi} 
+\sigma_{ge} e^{i\phi})/2$, for the local oscillator (LO) phase $\phi$, 
is measured by balanced homodyne detection (BHD) on the condition 
that the fluorescence intensity $I \propto \sigma_{ee}$ is measured at the 
detector $\mathrm{D}_I$. Assuming stationary dynamics, the normalized 
AIC function takes the following form 
\begin{eqnarray} 	\label{eq:htaudef}
h_{\phi}(\tau) = \frac{\langle: \sigma_{eg}(0) \sigma_{ge}(0) 
\sigma_{\phi}(\tau): \rangle }{ \alpha_{ee} \alpha_{\phi} } \,, 
\end{eqnarray}
where the steady state values of the intensity and the dipole quadrature 
amplitude are $\alpha_{ee}$ and 
$\alpha_{\phi} = \langle \sigma_{\phi} \rangle_{st}$, respectively, and the 
dots $::$ indicate normal and time operator ordering. For positive and 
negative time intervals we have 
\begin{subequations}
\begin{eqnarray} 
h_{\phi}(\tau \geq 0) &=& \frac{\langle \sigma_{eg}(0)  \sigma_{\phi}(\tau) 
	\sigma_{ge}(0) \rangle }{ \alpha_{ee} \alpha_{\phi} } \,, 		
	\label{eq:htaup} 		\\ 
h_{\phi}(\tau \leq 0) &=& \frac{ \mathrm{Re} \left[ 
	e^{-i \phi} \langle \sigma_{eg}(0)  \sigma_{ee}(-\tau) \rangle \right]}
	{ \alpha_{ee} \alpha_{\phi} } \,. 			\label{eq:htaun} 
\end{eqnarray}
\end{subequations}
For $\tau \geq 0$ a photon is detected at $\tau=0$, triggering the detection 
of a quadrature by balanced homodyne detection. For $\tau \leq 0$, on the 
other hand, it is the photon detection that follows the quadrature detection. 
Since the amplitude and intensity operators do not commute, there is no 
guarantee for symmetry to hold, that is, 
$h_{\phi}(- \tau) \neq h_{\phi}(\tau)$. The asymmetry results from the 
different fluctuations of the light's amplitude and intensity thanks to the 
breakdown of detailed balance in this system \cite{DeCC02}.  

%

We address this issue by analyzing noise properties of the fluorescence. 
We split the atomic operator dynamics into its mean plus fluctuations, 
$\sigma_{jk} = \alpha_{jk} +\Delta \sigma_{jk}$, where 
$\langle \Delta \sigma_{jk} \rangle =0$. The AIC correlation 
(\ref{eq:htaudef}) is split as \cite{hmcb10} 
\begin{eqnarray} 
h_{\phi}(\tau) = 1+ h_{\phi}^{(2)}(\tau) +h_{\phi}^{(3)}(\tau) \,, 	
\end{eqnarray}
where $h_{\phi}^{(2)}(\tau)$ and $h_{\phi}^{(3)}(\tau)$ are terms of second 
and third-order dipole fluctuations $\Delta \sigma_{jk}$ . For positive time 
intervals between photon and quadrature detection we have 
\begin{subequations} 	\label{eq:htauPf}
\begin{eqnarray} 	
h_{\phi}^{(2)}(\tau \geq 0) &=& \frac{ 2\mathrm{Re} [\alpha_{ge}  
\langle  \Delta \sigma_{eg}(0)  \Delta \sigma_{\phi}(\tau) \rangle]}
{ \alpha_{ee} \alpha_{\phi} } \,,     	\label{eq:htaup2} 	  \\ 
h_{\phi}^{(3)}(\tau \geq 0) &=& \frac{\langle \Delta \sigma_{eg}(0) 
\Delta \sigma_{\phi}(\tau) \Delta \sigma_{ge}(0) \rangle }
 { \alpha_{ee} \alpha_{\phi} } \,, 		\label{eq:htaup3}
\end{eqnarray}
\end{subequations}
where $\Delta \sigma_{\phi} =(\Delta \sigma_{eg} e^{-i\phi} 
+\Delta\sigma_{ge} e^{i\phi})/2$ is the dipole quadrature fluctuation 
operator.  For negative intervals, for which we reinforce notation with 
the superscript $(N)$, we have
\begin{eqnarray} 	\label{eq:htauNf}
h_{\phi}^{(N)} (\tau \leq 0) = 1+  \frac{ \mathrm{Re} [ e^{-i \phi} 
\langle  \Delta \sigma_{eg}(0)  \Delta \sigma_{ee}(-\tau) \rangle ] }
{\alpha_{ee} \alpha_{\phi} } \,.
\end{eqnarray}
This correlation is only of second-order in the dipole fluctuations. This 
differs from Eq.~(\ref{eq:htaup2}), however, by the presence of the 
time-dependent population noise operator $\Delta \sigma_{ee}$ instead 
of the time-dependent coherence fluctuation operators 
$\Delta \sigma_{eg}$ and $\Delta \sigma_{ge}$. 
%
\begin{figure}[t]
\includegraphics[width=8.5cm,height=8cm]{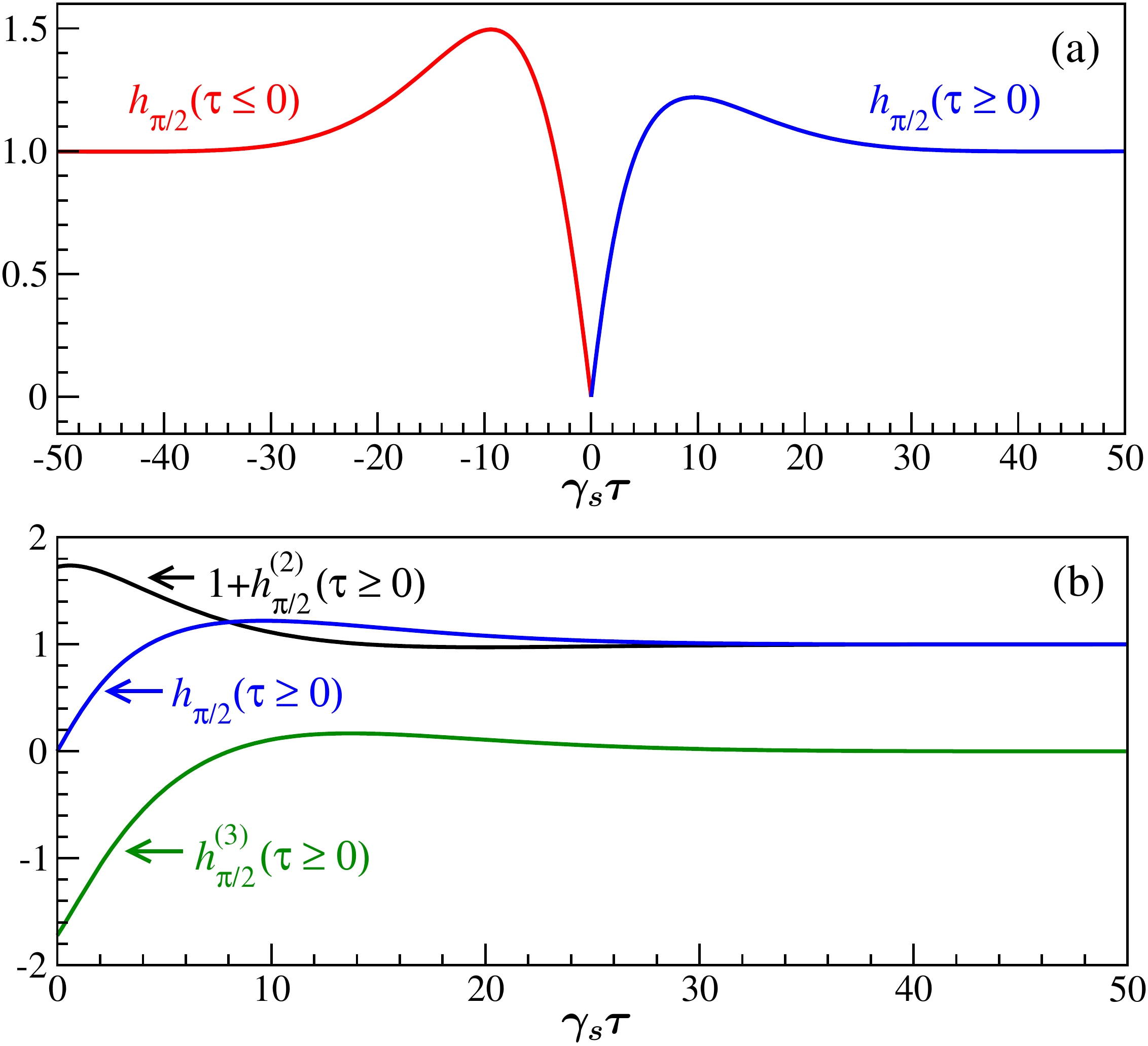}  
\vspace{2mm} 
\caption{\label{fig:chdW_satS} (a) Time-asymmetric AIC of light from 
the weak transition, for $\phi=\pi/2$. (b) Decomposition of 
$h_{\pi/2}(\tau \geq 0)$ (blue) into $1+h_{\pi/2}^{(2)}$ (black) and 
$h_{\pi/2}^{(3)}$ (green). The parameters used are: 
$\Omega_s =0.5 \gamma_s$, $\Omega_w =\gamma_w =0.1 \gamma_s$, 
$\Delta_s = \Delta_w =0$.}
\end{figure}

\begin{figure}[h]
\includegraphics[width=8.5cm,height=8cm]{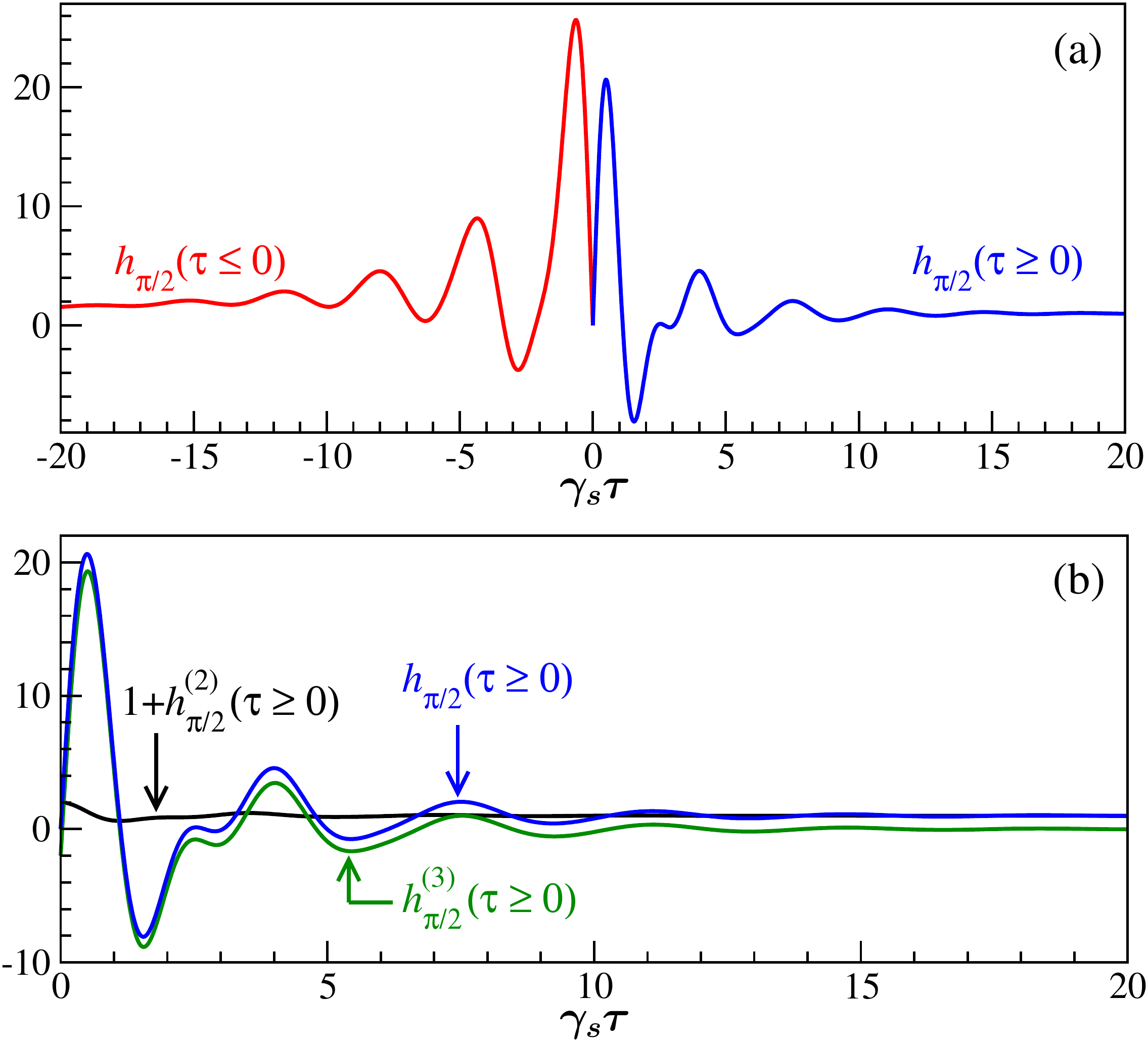}  
 \vspace{2mm} 
\caption{\label{fig:chdW_strongS} Same as in Fig.~\ref{fig:chdW_satS} 
except for the stronger driving of the strong transition, with 
$\Omega_s = 3.5 \gamma_s$. }
\end{figure}

Figures~\ref{fig:chdW_satS} and \ref{fig:chdW_strongS} show the AIC 
of the $\phi=\pi/2$ quadrature of the light from the weak transition for 
$\gamma_w=0.1 \gamma_s$. In Fig.~\ref{fig:chdW_satS} the Rabi 
frequency for the fast-decaying transition is moderately strong, 
$\Omega_s =0.5 \gamma_s$, and the Rabi frequency for the weak 
transition is $\Omega_w=0.1 \gamma_s$ (strong relative to 
$\gamma_w$). In Fig.~\ref{fig:chdW_satS}(a) the time asymmetry is 
evident. Fig.~\ref{fig:chdW_satS}(b)  shows the AIC decomposition for 
$\tau \geq 0$ where the second and third-order terms have similar size; 
this is a signature of a large deviation from Gaussian fluctuations. 

Figure~\ref{fig:chdW_strongS} shows the effect of the strong transition 
driven high above saturation, $\Omega_s = 3.5 \gamma_s$. In 
Figure~\ref{fig:chdW_strongS}(a) the asymmetry is still very clear and 
its size is increased compared to Fig.~\ref{fig:chdW_satS} due to the 
smaller values of $\alpha_{ww}$ and $\alpha_{\pi/2}$ used for the 
normalization; they are smaller because the system is in the regime of 
large quantum fluctuations due to the dressing of the ground state by the 
driving on the strong transition which detunes the weak transition laser 
by $\Omega_s/2$. As seen in the $g_{ww}^{(2)}$ photon correlation of 
Fig.~\ref{Fig_g2}(b), and in the frequency spectrum of 
Fig.~\ref{fig:specW_strongS}, these are the dominant frequencies of 
oscillations of $h_{\pi/2}(\tau)$ of the weak transition. There is a fast 
decay at $\sim 3 \gamma_s/4$, while the slow decay due to 
$\gamma_w$ can be seen only for long negative intervals. 
Figure~\ref{fig:chdW_strongS}(b) shows the dominance of the third-order 
term, $h_{\pi/2}(\tau \geq 0) \approx h_{\pi/2}^{(3)}(\tau \geq 0)$, that 
gives a strong signature of nonlinearity in the weak transition for 
$\Omega_w =\gamma_w$. It can be seen how well 
$h_{\pi/2}^{(3)}(\tau \geq 0)$ compares in size to 
$h_{\pi/2}^{(N)}(\tau \leq 0)$, which reflect population fluctuations. 

%

A comment regarding the AIC for the strong transition is in order, but it is 
not essential to show graphics here. It strongly resembles the cases of 
the two-level atom \cite{hmcb10} and of a different 3LA system 
\cite{CaRG16}. There is asymmetry, but very slight, and occurs in only 
two regimes: 1) just above saturation, $\Omega_s/\gamma_s \sim 1/4$, 
and 2) when the strong laser is detuned a few $\gamma_s$ from 
resonance. In the former the correlation looks like a more symmetric 
version of Fig.~\ref{fig:chdW_satS}. This is approximately the case in the 
AIC measurement of Ref.~\cite{GRS+09} for the strong transition of a 
$\Lambda$-type 3LA. In the latter the asymmetry is noticed only for long 
correlation time intervals, near the end of oscillations, as seen in Fig.~3(b) 
of Ref.~\cite{MaCa08}. In this case experimental background noise may 
hide the asymmetry. 

The AIC provides a strong assessment of the quantum nature of the 
emitted field through the violation of the classical inequalities 
\cite{CCFO00,FOCC00}: 
\begin{subequations} 	\label{eq:ineq}
\begin{eqnarray} 	
0 &\leq& h_{\phi} (\tau) - 1 \leq 1 	\,, 	\\
| h_{\phi} (\tau) - 1 | &\leq& | h_{\phi} (0) - 1 | \leq 1 \,.
\end{eqnarray}
\end{subequations}
Clearly, the AIC shown in Figs.~\ref{fig:chdW_satS} and 
\ref{fig:chdW_strongS} break several of these inequalities 
\cite{MaCa08,hmcb10,CaGH15}. One of them results from 
$h_{\pi/2}(0) =0$, which shows the emitted light's antibunching behavior;  
there is no field emitted when the atom is in the ground state. Thus, for 
short intervals $\tau$ the AIC is nonclassical. The other case is concerned 
with how large this correlation is by exceeding the classical bounds by 
orders of magnitude, as shown in Fig.~\ref{fig:chdW_strongS}, due to the 
small denominator in Eq.~\ref{eq:htaudef}. Its origin lies in the low photon 
emission rate that places the system in the regime of large quantum 
fluctuations.

As shown in Refs.~\cite{CCFO00,FOCC00}, the above classical bounds 
are stronger criteria for nonclassicality of the emitted field than squeezed 
light measurements, which provide a more familiar standard for probing 
phase-dependent fluctuations. We briefly consider this in the next Section. 
Detailed discussions on the hierarchy of measures of nonclassicality for 
higher-order correlation functions are presented in 
Refs.~\cite{ScVo05,ScVo06}. 

We close this Section with a discussion on the AIC for the $\phi=0$ 
quadrature. For nonzero detuning ($\Delta_w \neq 0$) the results are 
qualitatively similar to those of Figs.~\ref{fig:chdW_satS} and 
\ref{fig:chdW_strongS} but with a smaller amplitude. However, if 
$\Delta_w =0$, the mean dipole quadrature 
$\langle \sigma_0 \rangle_{st} = \alpha_0$ vanishes for all times, and 
so is the AIC (via the quantum regression formula). Thus, in order to have 
a nonzero signal, the dipole field must be mixed with a coherent offset 
before entering the detection setup \cite{CCFO00}. The resulting 
correlation, however, has a classical character, where features such as 
antibunching and the violation of classical inequalities, 
Eqs.~(\ref{eq:ineq}), are absent.

\section{Spectra and Variances of Quadratures}
In this Section we analyze noise properties in a quadrature of the emitted 
light field. The variance and the spectrum of squeezing had been the 
standard measures of quadrature fluctuations, hence it is convenient to 
include them in our analysis. More precisely, they are a natural part of the 
study of the AIC in the spectral domain 
\cite{CCFO00,FOCC00,hmcb10,CaGH15,CaRG16}. The AIC asymmetry 
and the large role of third-order fluctuations in resonance fluorescence 
clearly make the second-order correlation measurements for squeezing 
insufficient to explore most of the non-classical features of this and other 
quantum systems.  

\subsection{Spectral Fluctuations} 
\begin{figure}
\includegraphics[width=8.5cm,height=8cm]{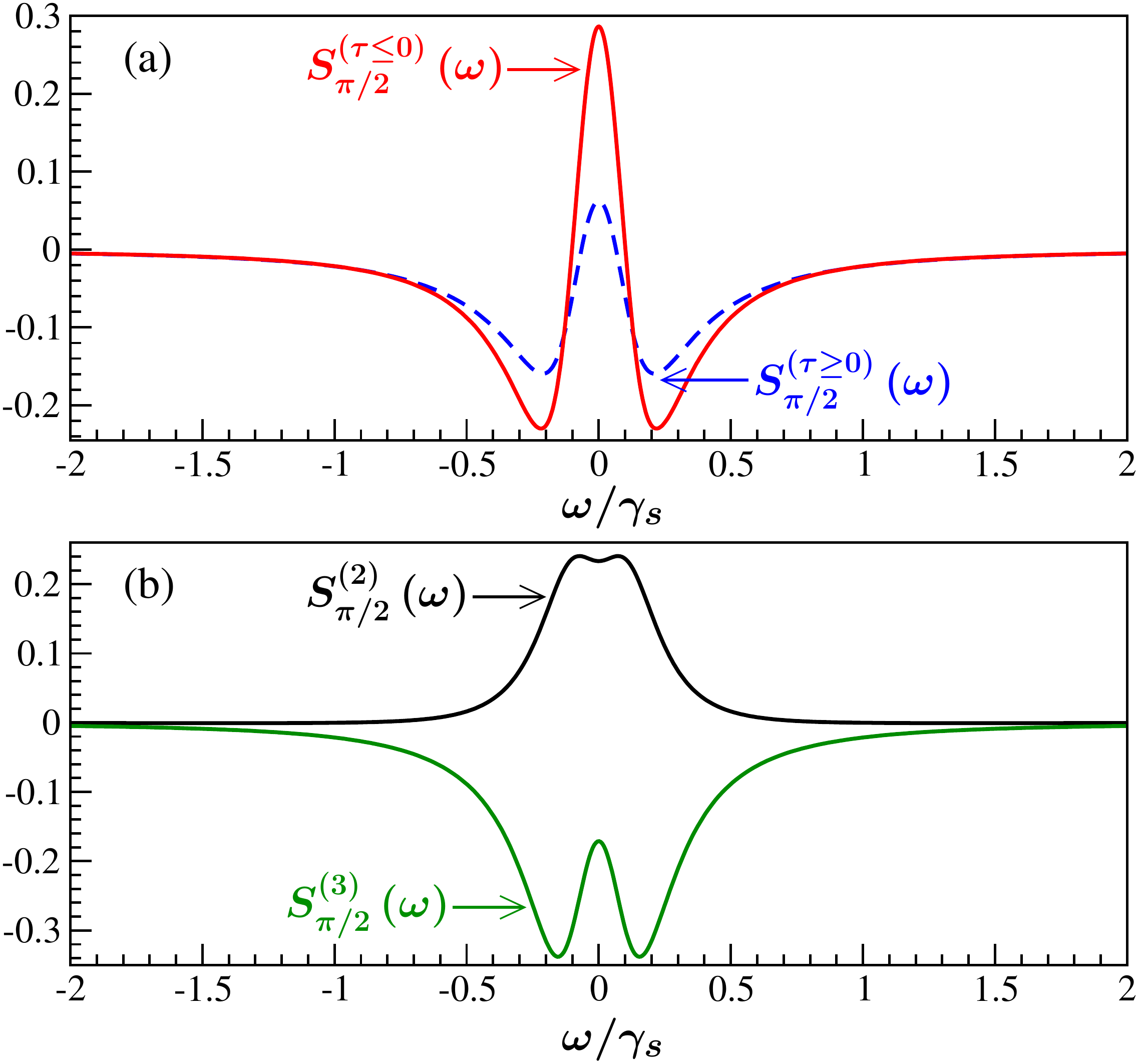}
\caption{\label{fig:specW_satS}
(a) Spectra, Eqs.~\ref{eq:chdspec}(a,b), for $\phi = \pi/2$ of the AIC 
of the weak transition plotted in Fig.~\ref{fig:chdW_satS}(a). (b) 
Decomposition of $S_{\pi/2}^{(\tau \geq 0)}$  into its terms of second 
and third order, Eqs.~\ref{eq:chdspec}(c). The parameters are: 
$\Omega_s =0.5 \gamma_s$, 
$\Omega_w =\gamma_w =0.1 \gamma_s$, $\Delta_s = \Delta_w =0$.}
\end{figure}
\begin{figure}[t]
\includegraphics[width=8.5cm,height=8cm]{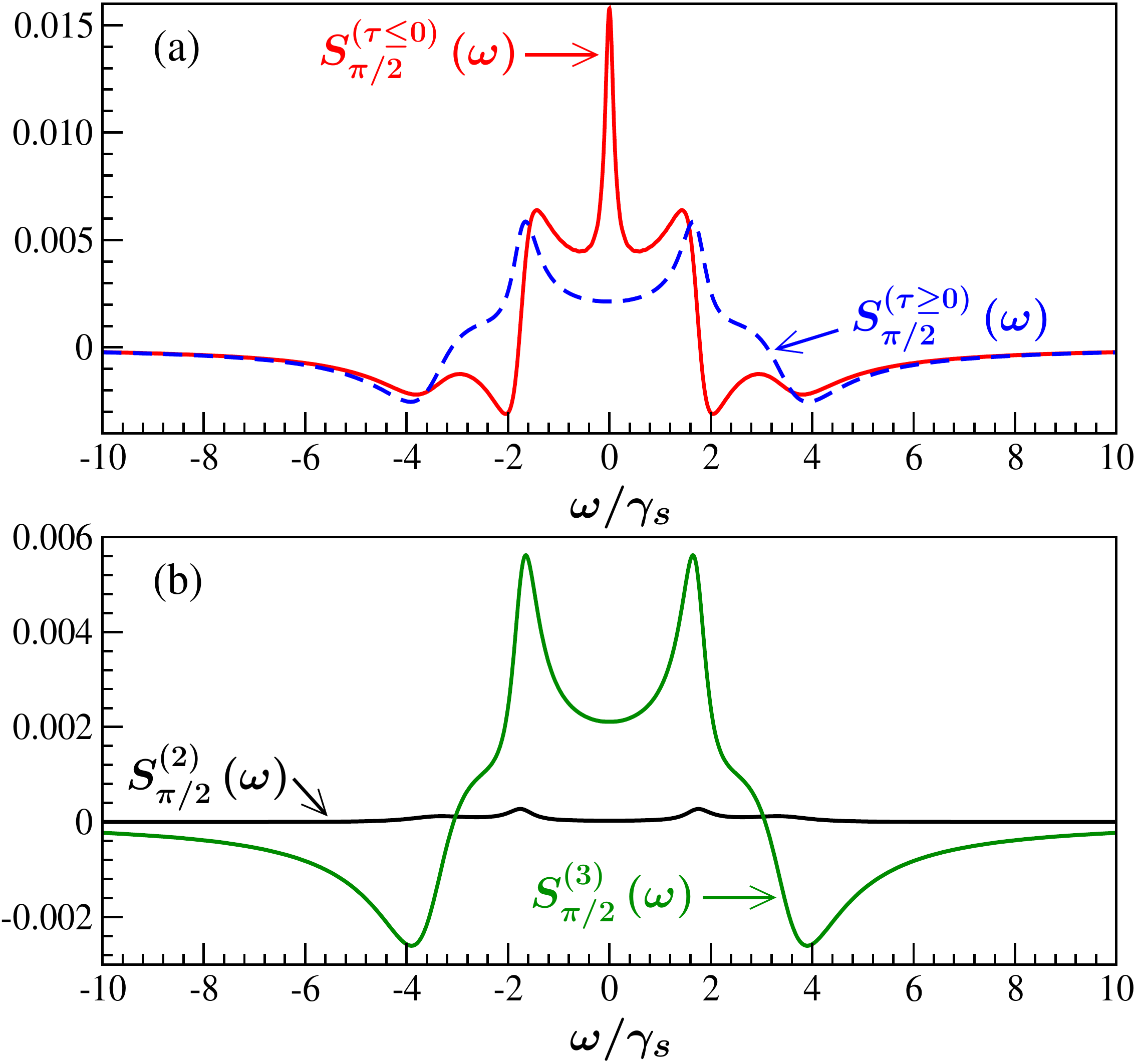}
\caption{\label{fig:specW_strongS}
(a) Spectra, Eqs.~\ref{eq:chdspec}(a,b), for $\phi = \pi/2$ of the AIC 
of the weak transition plotted in Fig.~\ref{fig:chdW_strongS}(a). (b) 
Decomposition of $S_{\pi/2}^{(\tau \geq 0)}$  into its terms of second 
and third order, Eqs.~\ref{eq:chdspec}(c). The parameters are: 
$\Omega_s =3.5 \gamma_s$, 
$\Omega_w =\gamma_w =0.1 \gamma_s$, $\Delta_s = \Delta_w =0$.}
\end{figure}
The spectral representation of fluctuations in the AIC provides 
complementary system information such as oscillation frequencies and 
decay rates and allows direct comparisons with measures of squeezed 
light. For the particular problem in this paper, the asymmetry of 
$h_{\phi}(\tau)$ would suggest the use of the Fourier exponential 
transform, 
$\int_{-\infty}^{\infty} \left[ h_{\phi}(\tau) -1 \right] e^{i \omega \tau} \,d\tau$. 
However, the fact that the AIC carries different information for positive 
and negative time intervals $\tau$, spectra should be obtained separately 
by the Fourier cosine transform,   
\begin{subequations} \label{eq:chdspec}
\begin{eqnarray} 	
S_{\phi}^{(\tau \leq 0)}(\omega) 
&=& 4\gamma_e \alpha_{ee} \int_{0}^{\infty}  
	\left[ h_{\phi} (\tau \leq 0) -1 \right] \cos{\omega \tau} \, d\tau   \,, \\ 
S_{\phi}^{(\tau \geq 0)}(\omega) 
&=& 4\gamma_e \alpha_{ee} \int_{0}^{\infty} 
	\left[ h_{\phi} (\tau \geq 0) -1 \right] \cos{\omega \tau} \, d\tau   \,, \\
&=& S_{\phi}^{(2)}(\omega) +S_{\phi}^{(3)}(\omega) 	\,,
\end{eqnarray}
\end{subequations}
where we took into account the splitting of $h_{\phi}(\tau \geq 0)$ into 
its second- and third-order terms. A Fourier exponential transform would 
only \textit{average} the spectra of both sides of $h_{\phi}(\tau)$, so 
incorrect information would be given. 

A signature of squeezed light is represented by the negative values of the 
frequency spectral function. Likewise, negative values in the AIC spectra 
indicate nonclassical light, beyond squeezing. It has been shown that the 
so-called \textit{spectrum of squeezing} \cite{CoWZ84,RiCa88} and the 
second-order spectrum are related as 
$S_{\phi}^{sq}(\omega) = \eta S_{\phi}^{(2)}(\omega)$ 
\cite{CCFO00,FOCC00}, where $\eta$ is a combined collection and 
detection efficiency. The AIC, due to its conditional detection nature, 
is independent of this $\eta$ factor. 

Figure~\ref{fig:specW_satS} shows the spectra of the AIC of 
Fig.~\ref{fig:chdW_satS}, with the strong transition excited moderately. 
The spectrum $S_{\pi/2}^{(\tau \geq 0)}(\omega)$ has a large central 
peak over a broad negative feature that reveals nonclassical features of 
the emitted light. Its decomposition, Fig.~\ref{fig:specW_satS}(b), shows 
that there is no squeezing, $S_{\pi/2}^{(2)}(\omega) \geq 0$, but a fully 
nonclassical (negative) bimodal feature is present in the third-order 
spectrum. The peaks are located around 
$\omega = \pm \Omega_s/2 = \pm 0.25 \gamma_s$. The spectrum 
$S_{\pi/2}^{(\tau \leq 0)}$ has a larger central peak, which reflects the 
larger amplitude of $h_{\pi/2}(\tau \leq 0)$. In both cases, the fact that 
the spectra have negative values reflects the presence of nonclassical 
effects such as antibunching and large fluctuations that lead to the 
violation of the inequalities in Eq.~(\ref{eq:ineq}). 

A stronger excitation of the $|g \rangle - |s \rangle$ transition makes it 
easier to extract spectral and transition dynamic information. 
Figure~\ref{fig:specW_strongS} gives the spectra of the AIC presented in 
Fig.~\ref{fig:chdW_strongS}. The peaks near 
$\pm \Omega_s/2 \,(\simeq \pm 1.75)$, due to the dressing of the strong 
transition, are reminiscent of the Autler-Townes effect, with slightly 
different splittings. For $S_{\pi/2}^{(\tau \leq 0)}$ in 
Fig.~\ref{fig:specW_strongS}(a), there is the outstanding feature of a 
narrow central peak, which reflects the slow decay at the rate 
$\gamma_w$, a remnant of electron shelving in the system 
\cite{CaRG16}. Such narrow peak is absent in $S_{\pi/2}^{(\tau \geq 0)}$, 
that is, there is no slow decay of $h_{\pi/2}(\tau \geq 0)$. The spectra 
$S_{\pi/2}^{(\tau \geq 0)}$ and $S_{\pi/2}^{(\tau \leq 0)}$ have a strong 
dispersive component because of the nonlinearity induced by driving the 
transition $|g \rangle - |w \rangle$ high above saturation. 
Fig.~\ref{fig:specW_strongS}(b) clearly shows the dominance of the 
third-order fluctuations \cite{hmcb10}. 
\begin{figure}
 \includegraphics[width=8.5cm,height=7cm]{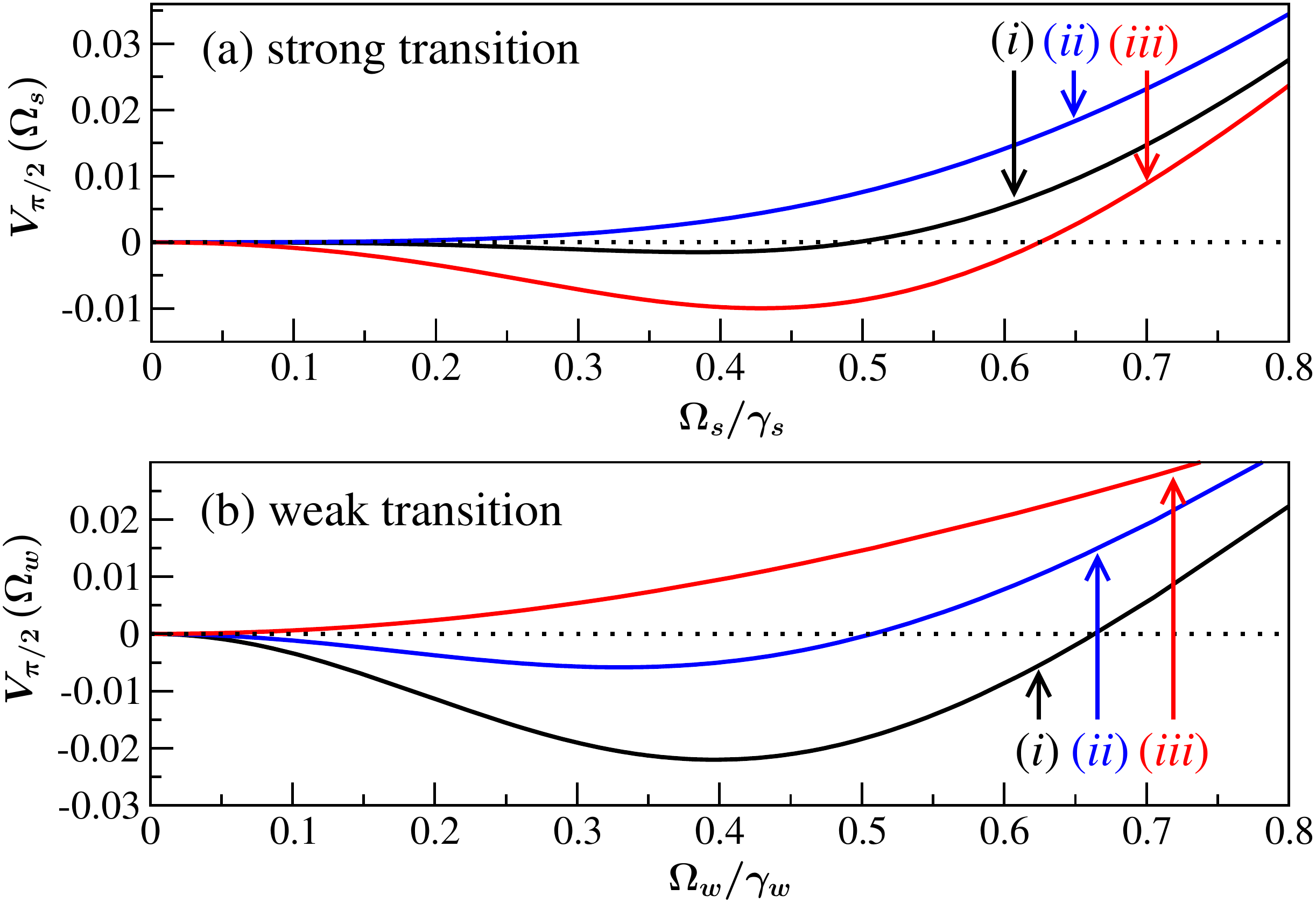}
\caption{\label{fig:varV3LA}
(a) Variance of fluorescence of the strong transition for $\phi=\pi/2$. 
The parameters used are: 
i) $\gamma_w = 0.01 \gamma_s$,    $\Omega_w =0.05 \gamma_s$, 
ii) $\gamma_w =0.01 \gamma_s$,   $\Omega_w =0.1 \gamma_s$, 
iii) $\gamma_w =0.1 \gamma_s$,     $\Omega_w =0.1 \gamma_s$.  
(b) Variance of fluorescence of the weak transition for $\phi=\pi/2$, 
$\gamma_w = 0.1 \gamma_s$, and: i) $\Omega_s=0.1 \gamma_s$, 
ii) $\Omega_s =0.2 \gamma_s$, iii) $\Omega_s =0.5 \gamma_s$. 
For all cases detunings are zero.}
\end{figure}

\subsection{Variance and Total Quadrature Noise} 
The noise in a quadrature is usually given by the variance 
\begin{equation} \label{eq:variance}
V_{\phi} = \langle : (\Delta \sigma_{\phi})^2  : \rangle 
	= \mathrm{Re} \left[ e^{-i \phi} 
	\langle \Delta \sigma_{eg} \Delta \sigma_{\phi} \rangle \right] 	\,,
\end{equation}
which is the unnormalized second-order amplitude-intensity correlation.  
The variance is related to the integrated spectrum of squeezing as  
$\int_{-\infty}{\infty} S_{\phi}^{(2)}(\omega) \,d\omega 
= 4\pi \gamma_e \eta V_{\phi}$. Negative values of the variance are a 
signature of squeezed fluctuations. 
For the strong transition, the squeezing for $\phi=\pi/2$ is small or null, 
Fig.~\ref{fig:varV3LA}(a), compared to the case of a two-level atom  \cite{RiCa88,CaRG16,SHJ+15}. The reduction of squeezing is due to the 
added incoherent emission in the weak transition. The weak transition 
also features squeezing, Fig.~\ref{fig:varV3LA}(b), but not much larger 
than for the strong transition.
%
\begin{figure}[t]
 \includegraphics[width=8.5cm,height=7cm]{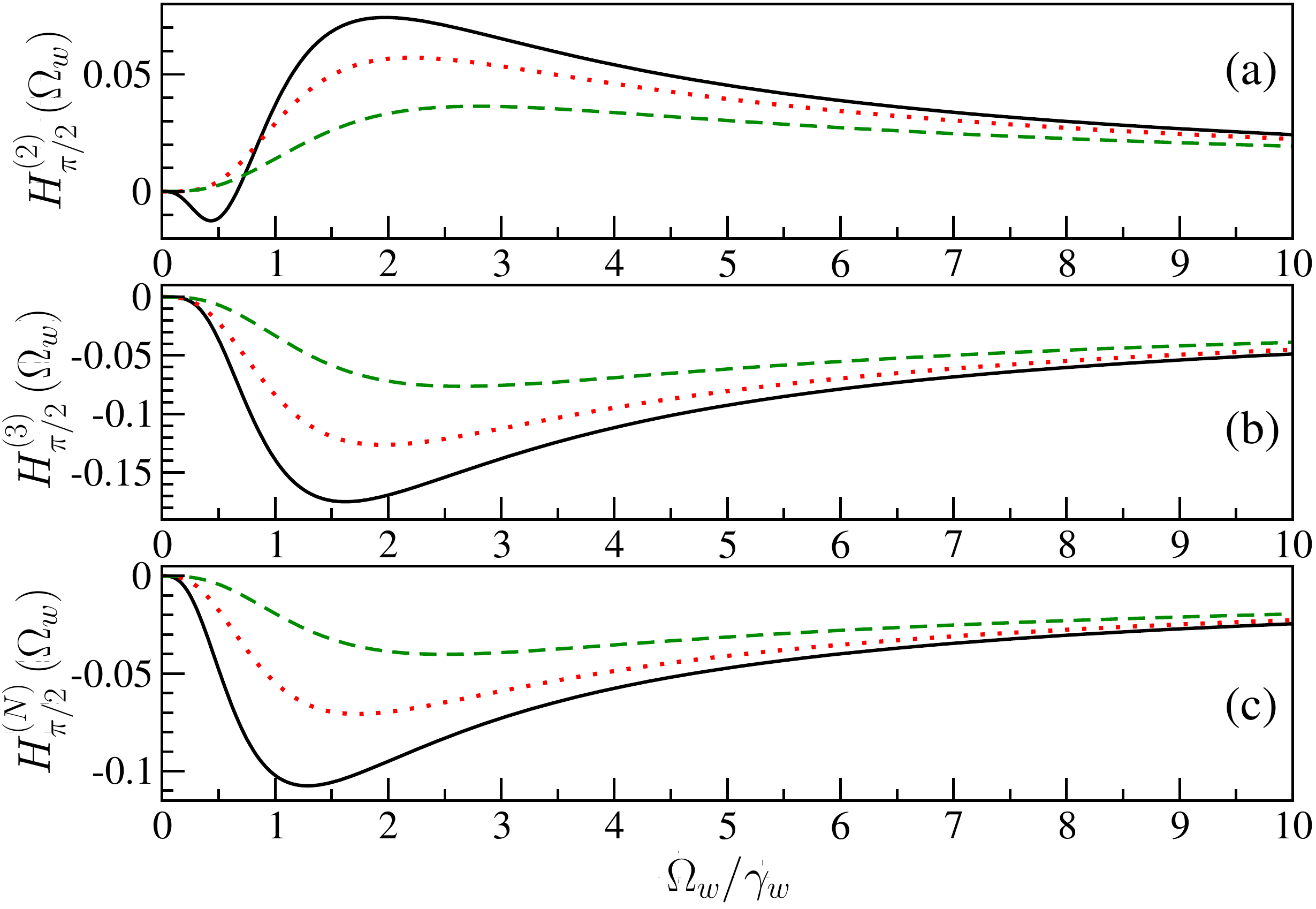}
\caption{\label{fig:noiseW}
Noise for $\phi = \pi/2$ from CHD of the weak transition in (a) second 
order, (b) third order for $\tau \geq 0$, and (c) for $\tau \leq 0$. 
Parameters are: $\gamma_w = 0.1 \gamma_s$,
$\Omega_s =0.1 \gamma_s$ (solid-black), 
$\Omega_s =0.5 \gamma_s$ (dotted-red), 
$\Omega_s =0.9 \gamma_s$ (dashed-green). 
For all curves detunings are zero. }
\end{figure}

As for the spectrum of squeezing, the variance is not the right measure 
of noise for the AIC. It has two problems. The first is the asymmetry itself, 
which forces us to consider the noise from the positive and the negative 
time interval parts of the correlation separately. The second is that it is 
not possible to separate experimentally the second- and third-order terms 
in the $\tau \geq 0$ part. We have, however, found it useful to perform 
such separation in order to explain the origin of nonclassical features of 
the AIC correlation functions. A natural choice to measure the noise is to 
integrate the spectra of Figs.~\ref{fig:specW_satS} and 
\ref{fig:specW_strongS}, which are proportional to the initial values of the unnormalized correlations (\ref{eq:htauPf}) and (\ref{eq:htauNf}), 
\begin{subequations} 	\label{eq:noise}
\begin{eqnarray} 	
H_{\phi}^{(2)} &=& 2 \mathrm{Re} [\alpha_{ge}  
	\langle  \Delta \sigma_{eg}  \Delta \sigma_{\phi} \rangle] \,,     		\label{eq:noise2} 	  \\ 
H_{\phi}^{(3)} &=& \langle \Delta \sigma_{eg} 
	\Delta \sigma_{\phi} \Delta \sigma_{ge} \rangle \,, 		
	\label{eq:noise3} 	\\
H_{\phi}^{(N)} &=& \mathrm{Re} [ e^{-i \phi} \langle  
	\Delta \sigma_{eg}  \Delta \sigma_{ee} \rangle ]  \,,  
	\label{eq:noiseN}
\end{eqnarray}
\end{subequations} 
where the noise correlations $\langle \cdots \rangle$ are given in Eqs. 
(\ref{eq:tauzerocorrel}). 

In Fig.~\ref{fig:noiseW} we plot such noise functions of the $\phi =\pi/2$ 
quadrature of the weak transition. Since the variance is proportional to 
$H_{\phi}^{(2)}$, there is a narrow range of Rabi frequencies $\Omega_w$ 
where the noise is negative [compare line (i) in Fig.~\ref{fig:varV3LA}(b) 
and solid line in Fig.~\ref{fig:noiseW}(a)]. Stronger driving in both 
transitions destroys squeezing. The third-order term is seen to be 
nonclassical for a wider range of Rabi frequencies, Fig.~\ref{fig:noiseW}(b), 
and it is the dominant term of the total noise. Since these noise functions 
are evaluated at $\tau =0$, we obtain 
$H_{\phi}^{(N)}(0) = H_{\phi}^{(2)}(0) +H_{\phi}^{(3)}(0)$, which can be 
readily seen in Fig.~\ref{fig:noiseW}(c).  

\section{Conclusions}
\vspace{-0.05cm}
We have investigated quantum fluctuations of the scattered light from  
two monochromatically driven transitions of a V-type 3LA, mainly using 
phase-dependent amplitude-intensity correlations. The noncommutativity 
of amplitude and intensity operators and the competition among transitions 
make the correlation asymmetric and large, more pronounced for the weak  
transition, which has larger quantum fluctuations than the strong one. The 
asymmetry is a signature of nonclassical and non-Gaussian fluctuations 
(hence the breakdown of detailed balance) which manifest as different 
oscillation frequencies and decays in the spectra (and integrated spectra) 
associated with the positive and negative time intervals of the AIC 
correlations. Third-order fluctuations, with their dispersive spectral shape, 
are dominant for strong excitation of a transition. We also showed that the 
correlation of two photons from the weak transition has enhanced 
nonclassical behavior of intensity fluctuations: even for strong excitation, 
the correlation evolves most of the time below the level of a coherent state. 
CHD is a valuable tool to study the AIC to reveal not only nonclassical 
states of the field fluctuations but also to explore nonequilibrium physics in microscopic systems. 
\vspace{-0.5cm}
\section*{Acknowledgments}
HMCB thanks support from Project CONACYT-FOMIX, M\'exico, 
No. 225447; RRA thanks CONACYT, M\'exico, for Scholarship 
No. 379732 and DGAPA-UNAM, M\'exico, for support under Project 
No. IN113016.  
\section*{Appendix}
To calculate correlations, spectra, variances, and noise, we evaluate the 
zero interval, steady state, correlations:
\begin{subequations} 	\label{eq:tauzerocorrel}
\begin{eqnarray} 
\langle \Delta \sigma_{eg} \Delta \sigma_{ge} \rangle
	&=& \alpha_{ee}  - |\alpha_{eg}|^2 \,,	\\ 
\langle \Delta \sigma_{eg} \Delta \sigma_{eg} \rangle 
	&=&  - \alpha_{eg}^2 \,, 	\\
\langle \Delta \sigma_{eg} \Delta \sigma_{ee} \rangle
	&=& -\alpha_{eg}  \alpha_{ee} \,, 	\\
\langle \Delta \sigma_{eg} \Delta \sigma_{ge} \Delta \sigma_{ge} \rangle
	&=& 2\alpha_{ge} ( |\alpha_{eg}|^2 -\alpha_{ee} )  	\,, \\ 
\langle \Delta \sigma_{eg} \Delta \sigma_{eg} \Delta \sigma_{ge} \rangle
	&=& 2\alpha_{eg} ( |\alpha_{eg}|^2 -\alpha_{ee} ) 	\,, 
\end{eqnarray}
\end{subequations}
where $\Delta \sigma_{jk} = \sigma_{jk} - \alpha_{jk}$. 

Additionally, we observe that $h_{\phi}(0)=0$ (as for antibunching, given 
the fermionic character of the dipole operators). From 
Eqs.~(\ref{eq:htauPf}) and (\ref{eq:tauzerocorrel}), we can calculate the 
third-order AIC correlation for $\tau =0$,  
\begin{eqnarray} 
h_{\phi}^{(3)}(0) = -\left[ 1+h_{\phi}^{(2)}(0) \right] 
	= \frac{2(|\alpha_{eg}|^2 -\alpha_{ee})}{\alpha_{ee}} 	\,. 
\end{eqnarray}
For strong driving $h_{\phi}^{(3)}(0) \to -2$, whether it is the strong 
transition or the weak one. In this regime, the stationary excited state 
population is bound as $\alpha_{ee} < 0.5$, while the 
coherence is very small,  $|\alpha_{eg}|^2 \sim \Omega_e^{-2} \to 0$.


\end{document}